# Predicting equation from Chaotic data by Nonlinear Singular Value Decomposition

( Prabhakar G.Vaidya & P.S. Sajini Anand)

*Mathematical Modeling Unit*
*National Institute of Advanced Studies*
*Indian Institute of Science Campus*
*Bangalore - 560012*
*Phone : +91 80 23602050*
*email: sajini@nias.iisc.ernet.in*

## Abstract

Singular Value Decomposition can be considered as an effective method for Signal Processing/especially data compression. In this short paper we investigate the application of SVD to predict data equation from data. The method is similar to nonlinear ARMA method for fitting a nonlinear equation to the data.

## Introduction

Von Neumann, during World War II came out with a data, which looked random. Frequency analysis of data showed that it has a wide spectrum and poor correlation. So many came up with a conclusion that the data is random. But the data was generated by a simple dynamical system the Logistic Map: $X_{n+1} = 4 \cdot X_n \cdot (1 - X_n)$

And now it is well known that a deterministic dynamical system can generate random looking or chaotic sequences. We know that practical data in most cases are non-stationary, it may include noise and it is always generated by some non-linear mechanisms. The data can appear complex, if there is some non-linearity. It is also possible that the underlying dynamics might be of lower dimensions.

## Geometry of chaotic data

State space reconstruction of data is the creation of a higher dimensional, deterministic state space from a lower dimensional time series. This is an old idea in statistic literature [Whitney 1936]. Later this idea was introduced to dynamical systems [Packard et. al. 1980, Aeyels D. 1980, Takens 1981]. This method became very important in dynamical systems because it was demonstrated that the reconstructed state spaces do preserve geometrical invariants like eigen values of a fixed point, fractal dimension of attractor, Lyapunov exponent of trajectory, topological entropy etc.

**Takens Embedding**

Given the time series data a simple embedding can be represented in terms of a set of vectors which can be arranged in the form of a matrix as shown below.

$$\begin{bmatrix} x(t) & x(t+\tau) & x(t+2\tau) & \cdots & \cdots & x[t+(n-m)\tau] \\ x(t+\tau) & x(t+2\tau) & x(t+3\tau) & \cdots & \cdots & \vdots \\ x(t+2\tau) & x(t+3\tau) & x(t+4\tau) & \cdots & \cdots & \vdots \\ \vdots & \vdots & \vdots & \vdots & \vdots & \vdots \\ \vdots & \vdots & \vdots & \vdots & \vdots & \vdots \\ x[t+(m-1)\tau] & \cdots & \cdots & \cdots & \cdots & x[t+(n-1)\tau] \end{bmatrix}$$

This matrix is a collection of vectors each of dimension "m". According to the Whitney embedding theorem, the time delay state space reconstruction is an embedding if m>2D.

In the time delay matrix, each column is regarded as a vector or point in a higher dimensions and the dynamics of the system is supposed to proceed from column to column.

**Fitting a polynomial to data**

Suppose some data is available from a nonlinear system. For example if the data is generated from a simple dynamical system like the logistic map. Now the data is going to look like chaotic/random. It is possible to fit a higher order polynomial to the data.
………………….(merits and demerits of fitting a polynomial)

**SVD and embedding**

Here we describe a single method to fit a nonlinear equation to the data with the help of singular value decomposition and time delayed embedding.

We know that the original dynamics of logistic map is only one-dimensional, even if we embed the data in 3 or higher dimensions it has still one-dimensional dynamics. That is the projection of manifold in to 2D looks like a parabola. i.e. when we do embedding , this one dimensional manifold is sitting in a higher dimensional object say $R^n$.

We have taken the data, made the delayed vectors and created the matrix out of those vectors. i.e. we have embedded it in a higher dimension 'n'.
According to Whitney's theorem we need an embedding dimension greater than 2.
For practical purpose let us set 'n' as 4 or 5.

Now we need to add some nonlinear columns to the delayed matrix. It can be the square of any column or the product of 2 columns and so on, if you need to predict a quadratic equation. It can be the cube of any column or the product of columns if you need predict a cubic equation.

If you do the Singular Value Decomposition of the extended time delayed matrix,

$$N = U \cdot W \cdot V^T$$

Examine the orthogonal matrices U and V and the singular values of N. If there is any relation between the time delayed vectors and corresponding nonlinear columns, corresponding singular value will go to zero. We can extract information from the vector in V matrix, which corresponds to the nonlinear columns of N.

## Programming Details

SVD nonlinear ARMA model
How to predict equations from data? Suppose some data is available from a nonlinear system. You have given the data. You need to predict the equation from data. Here data is generated from a simple dynamical system the logistic map.

## Generation of data

$h := .01$

$$X := \left| \begin{array}{l} X_0 \leftarrow 0 \\ \text{for } n \in 1..100 \\ \quad X_n \leftarrow X_{n-1} + h \\ \text{for } n \in 1..100 \\ \quad Y_n \leftarrow 4 \cdot X_n \cdot (1 - X_n) \\ (X \;\; Y) \end{array} \right.$$

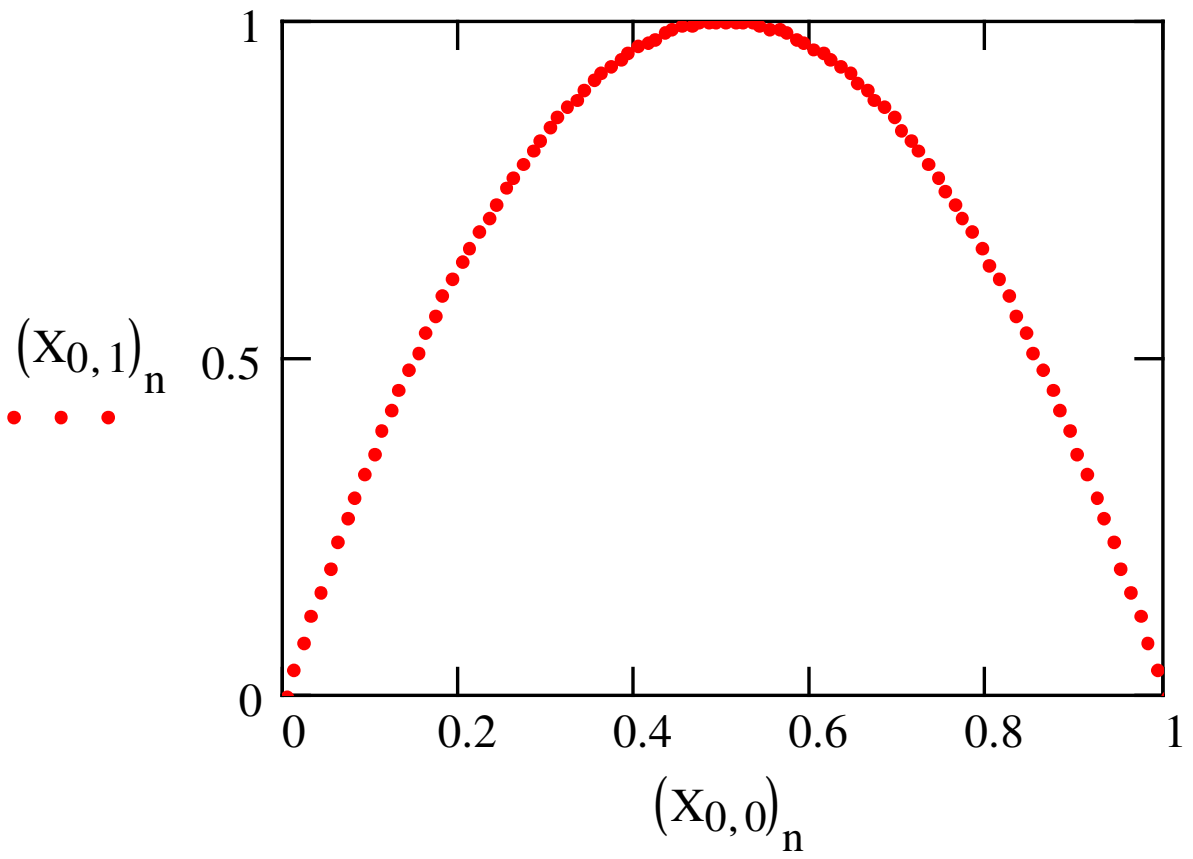

$n := 0..100$

$Y := (X_{0,1})$

$g(x) := 4 \cdot x \cdot (1 - x)$ Function is defined

We have selected an initial condition and applied map 1000 times and generated data z
Initial condition, $z_0 := .02$

$$z := \left| \begin{array}{l} z_0 \leftarrow z0 \\ \text{for } n \in 1..1000 \\ \quad z_n \leftarrow g(z_{n-1}) \\ z \end{array} \right.$$

$$n := 0..1000$$

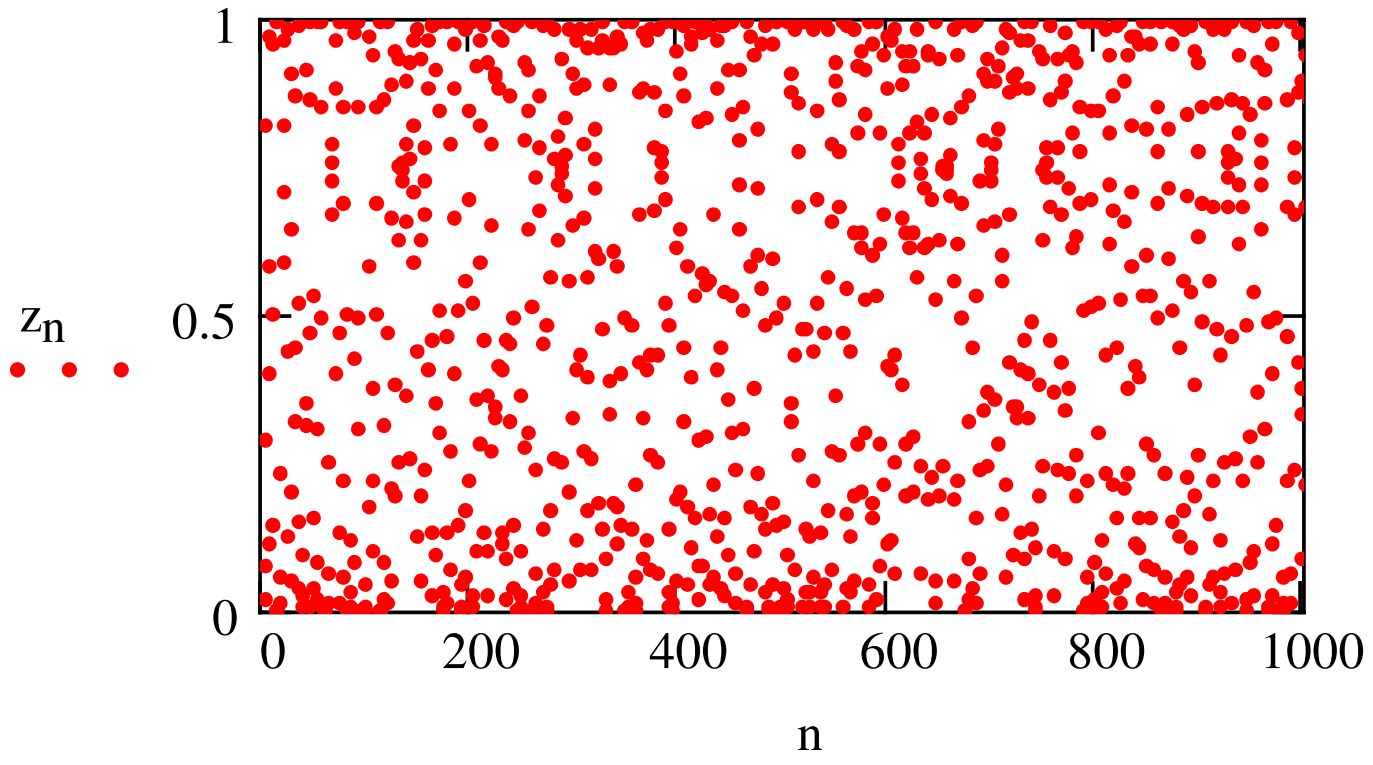

So the graph shows the randomly spaced data points generated from the chaotic logistic map. Plotting z versus z+1 we get the logistic map (time delayed embedding in 2D)

z =

| | 0 |
|---|---|
| 0 | 0.02 |
| 1 | 0.0784 |
| 2 | 0.2890138 |
| 3 | 0.8219392 |
| 4 | 0.5854205 |
| 5 | 0.9708133 |
| 6 | 0.1133392 |
| 7 | 0.4019738 |
| 8 | 0.9615635 |
| 9 | 0.1478366 |
| 10 | 0.5039236 |
| 11 | 0.9999384 |
| 12 | 0.0002463 |
| 13 | 0.000985 |
| 14 | 0.003936 |
| 15 | 0.0156821 |

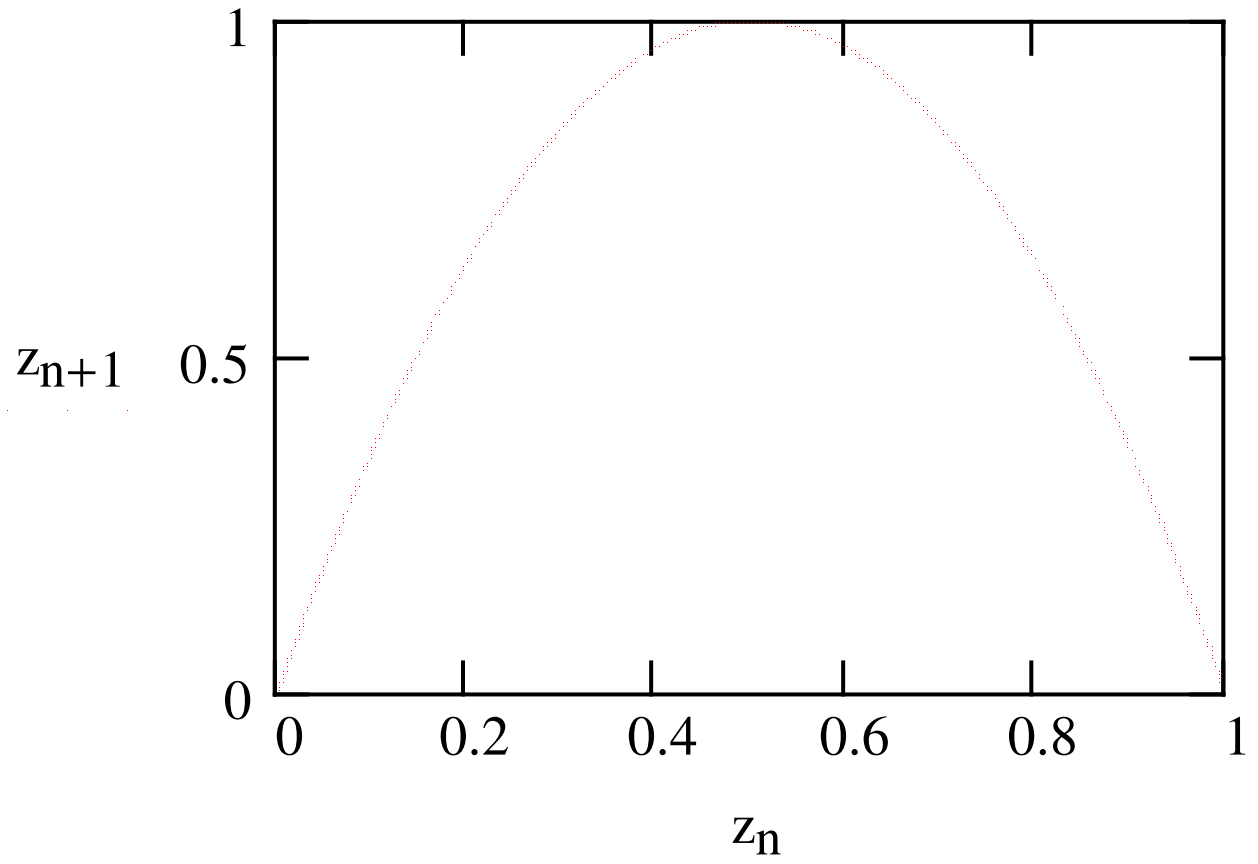

## Prediction of Equation from Data

### (1) Prediction of Quadratic Equation

Now our aim is to predict the equation from the data points.
First of consider the time delayed vectors, when m = 5 for example a 5x 5 matrix of delayed vectors is shown below

$$\begin{pmatrix} z_0 & z_1 & z_2 & z_3 & z_4 \\ z_1 & z_2 & z_3 & z_4 & z_5 \\ z_2 & z_3 & z_4 & z_5 & z_6 \\ z_4 & z_5 & z_6 & z_7 & z_8 \\ z_5 & z_6 & z_7 & z_8 & z_9 \end{pmatrix}$$

This program creates Takens matrix for any number of columns and rows

```
N := | z ← z
     | k ← 0
     | for i ∈ k .. k + 5
     |     | for j ∈ k .. k + 4
     |     |    y[i, j−k] ← z[j]
     |     | k ← k + 1
     | y
```

$$N = \begin{pmatrix} 0.02 & 0.0784 & 0.2890138 & 0.8219392 & 0.5854205 \\ 0.0784 & 0.2890138 & 0.8219392 & 0.5854205 & 0.9708133 \\ 0.2890138 & 0.8219392 & 0.5854205 & 0.9708133 & 0.1133392 \\ 0.8219392 & 0.5854205 & 0.9708133 & 0.1133392 & 0.4019738 \\ 0.5854205 & 0.9708133 & 0.1133392 & 0.4019738 & 0.9615635 \\ 0.9708133 & 0.1133392 & 0.4019738 & 0.9615635 & 0.1478366 \end{pmatrix}$$

Now let us add the square first column to the matrix.

```
N := | N ← N
     | for i ∈ 0 .. 5
     |    N[i, 5] ← (N[i, 0])^2
     | N
```

Now lets do the singular value decomposition of matrix N

$$N = \begin{pmatrix} 0.02 & 0.0784 & 0.2890138 & 0.8219392 & 0.5854205 & 0.0004 \\ 0.0784 & 0.2890138 & 0.8219392 & 0.5854205 & 0.9708133 & 0.0061466 \\ 0.2890138 & 0.8219392 & 0.5854205 & 0.9708133 & 0.1133392 & 0.083529 \\ 0.8219392 & 0.5854205 & 0.9708133 & 0.1133392 & 0.4019738 & 0.6755841 \\ 0.5854205 & 0.9708133 & 0.1133392 & 0.4019738 & 0.9615635 & 0.3427172 \\ 0.9708133 & 0.1133392 & 0.4019738 & 0.9615635 & 0.1478366 & 0.9424785 \end{pmatrix}$$

m := rows(N)

n := cols(N)

U := submatrix(svd(N), 0, m − 1, 0, n − 1)

$N = U \cdot W \cdot V^T$

V := submatrix(svd(N), m, m + n − 1, 0, n − 1)

W := diag(svds(N))

$$U \cdot W \cdot V^T = \begin{pmatrix} 0.02 & 0.0784 & 0.2890138 & 0.8219392 & 0.5854205 & 0.0004 \\ 0.0784 & 0.2890138 & 0.8219392 & 0.5854205 & 0.9708133 & 0.0061466 \\ 0.2890138 & 0.8219392 & 0.5854205 & 0.9708133 & 0.1133392 & 0.083529 \\ 0.8219392 & 0.5854205 & 0.9708133 & 0.1133392 & 0.4019738 & 0.6755841 \\ 0.5854205 & 0.9708133 & 0.1133392 & 0.4019738 & 0.9615635 & 0.3427172 \\ 0.9708133 & 0.1133392 & 0.4019738 & 0.9615635 & 0.1478366 & 0.9424785 \end{pmatrix}$$

$$U = \begin{pmatrix} -0.2620256 & -0.3675629 & 0.4342151 & -0.0656394 & 0.2846087 & -0.7227633 \\ -0.3845613 & -0.5330007 & -0.0017317 & -0.5333821 & 0.1410949 & 0.5134351 \\ -0.4035742 & -0.140452 & 0.336676 & 0.3084613 & -0.7755 & 0.0866124 \\ -0.4562703 & 0.3449357 & -0.5070991 & -0.4489346 & -0.266035 & -0.3786429 \\ -0.4433201 & -0.2090544 & -0.4989506 & 0.6426856 & 0.3110953 & 0.0314141 \\ -0.4646158 & 0.6311867 & 0.4381808 & 0.0382045 & 0.3607421 & 0.2492763 \end{pmatrix}$$

$$V = \begin{pmatrix} -0.4030338 & 0.5369522 & -0.1741709 & 0.1122791 & 0.1464768 & -0.6963106 \\ -0.3951893 & -0.1787735 & -0.4130562 & 0.5712263 & -0.5335237 & 0.1740777 \\ -0.4259591 & -0.0483877 & -0.0505329 & -0.7759607 & -0.4599445 & 0 \\ -0.4913916 & -0.1479483 & 0.8294358 & 0.2205122 & 0.0074986 & 0 \\ -0.4073358 & -0.5631879 & -0.3217203 & -0.0970639 & 0.6355875 & 0 \\ -0.3042365 & 0.5816456 & -0.0709068 & -0.0305274 & 0.2798577 & 0.6963106 \end{pmatrix}$$

$$W = \begin{pmatrix} 3.0708065 & 0 & 0 & 0 & 0 & 0 \\ 0 & 1.2741632 & 0 & 0 & 0 & 0 \\ 0 & 0 & 1.0200109 & 0 & 0 & 0 \\ 0 & 0 & 0 & 0.804721 & 0 & 0 \\ 0 & 0 & 0 & 0 & 0.7256714 & 0 \\ 0 & 0 & 0 & 0 & 0 & 0 \end{pmatrix}$$

Notice that the V matrix has its last column with some zero elements in it.

$$N = U \cdot W \cdot V^T$$

$$U^T \cdot N \cdot V = W$$

Consider matrix U. now u (n) is defined as the transpose of nth column of U.

$$u(n) := U^{\langle n \rangle T}$$

Similarly v (n) is the nth column of V

$$v(n) := V^{\langle n \rangle}$$

So the nth singular value is given by

$$\lambda(n) := u(n) \cdot N \cdot v(n)$$

And the last singular value is zero

$$\lambda(5) = (u(5) \cdot N \cdot v(5)) = 0$$

$$u(5) = \begin{pmatrix} -0.7227633 & 0.5134351 & 0.0866124 & -0.3786429 & 0.0314141 & 0.2492763 \end{pmatrix}$$

$$v(5) = \begin{pmatrix} -0.6963106 \\ 0.1740777 \\ 0 \\ 0 \\ 0 \\ 0.6963106 \end{pmatrix}$$

now

$$\lambda(5) = (u(5) \cdot N \cdot v(5)) = 0$$

$$u(5) \cdot N = \begin{pmatrix} 0 & 0 & 0 & 0 & 0 & 0 \end{pmatrix}$$

$$N \cdot v(5) = \begin{pmatrix} 0 \\ 0 \\ 0 \\ 0 \\ 0 \\ 0 \end{pmatrix}$$

$$N^{\langle 0 \rangle} \cdot v(5)_0 + N^{\langle 1 \rangle} \cdot v(5)_1 + N^{\langle 2 \rangle} \cdot v(5)_2 + N^{\langle 3 \rangle} \cdot v(5)_3 + N^{\langle 4 \rangle} \cdot v(5)_4 + N^{\langle 5 \rangle} \cdot v(5)_5 = \begin{pmatrix} 0 \\ 0 \\ 0 \\ 0 \\ 0 \\ 0 \end{pmatrix}$$

$$N^{\langle 0 \rangle} \cdot v(5)_0 + N^{\langle 1 \rangle} \cdot v(5)_1 + N^{\langle 5 \rangle} \cdot v(5)_5 = \begin{pmatrix} 0 \\ 0 \\ 0 \\ 0 \\ 0 \\ 0 \end{pmatrix}$$

First column in the time delay matrix is Xn

$$N^{\langle 0 \rangle} = \begin{pmatrix} 0.02 \\ 0.0784 \\ 0.2890138 \\ 0.8219392 \\ 0.5854205 \\ 0.9708133 \end{pmatrix}$$

Second column in the time delay matrix is Xn+1

$$N^{\langle 1\rangle} = \begin{pmatrix} 0.0784 \\ 0.2890138 \\ 0.8219392 \\ 0.5854205 \\ 0.9708133 \\ 0.1133392 \end{pmatrix}$$

The last column is Xn square

$$N^{\langle 5\rangle} = \begin{pmatrix} 0.0004 \\ 0.0061466 \\ 0.083529 \\ 0.6755841 \\ 0.3427172 \\ 0.9424785 \end{pmatrix}$$

$$N^{\langle 0\rangle} \cdot v(5)_0 + N^{\langle 1\rangle} \cdot v(5)_1 + N^{\langle 5\rangle} \cdot v(5)_5 = \begin{pmatrix} 0 \\ 0 \\ 0 \\ 0 \\ 0 \\ 0 \end{pmatrix}$$

We know that

$v(5)_0 = -0.6963106$

$v(5)_1 = 0.1740777$

$v(5)_5 = 0.6963106$

So solving the equation

$$-0.6963106 \cdot X_n + 0.1740777 \cdot X_{n+1} + 0.6963106 \cdot (X_n)^2 = 0$$

$$0.6963106 \cdot X_n - 0.6963106 \cdot (X_n)^2 = 0.1740777 \cdot X_{n+1}$$

$$\frac{0.6963106}{0.1740777} \cdot X_n + \frac{-0.6963106}{0.1740777} \cdot (X_n)^2 = X_{n+1}$$

and

$$\frac{0.6963106}{0.1740777} = 4$$

We get

$$4 \cdot X_n - 4 \cdot (X_n)^2 = X_{n+1}$$

$$4 \cdot X_n \cdot (1 - X_n) = X_{n+1}$$

**Conclusion**

The non-linear Quadratic equation is predicted from data. The method works for cubic and quartic cases too.